# Structure and $T_c$ of $Y_{0.8}Ca_{0.2}Ba_2Cu_3O_{6.92}$ under High Pressure.

**Walter H. Fietz, Frank W. Hornung, Kai Grube, Sonja I. Schlachter, Thomas Wolf, Bernhard Obst, and Peter Schweiss***

*Forschungszentrum Karlsruhe, Institut für Technische Physik and*
*\* Institut für Nukleare Festkörperphysik*
*Postfach 3640, 76021 Karlsruhe, Germany*

*The lattice parameters and $T_c$ of $Y_{1-y}Ca_yBa_2Cu_3O_x$ have been determined under pressure up to 10 GPa using a diamond anvil cell. The results are used to test the extended charge transfer model that connects the pressure effect on $T_c$ to an intrinsic $dT_{c,max}/dp$ and a pressure induced hole doping $dn_h/dp$, the latter being caused by charge transfer from the CuO chains to the $CuO_2$ planes. $dn_h/dp$ and $dT_{c,max}/dp$ are usually assumed to be constant with respect to pressure p. However, our experiments show that the usage of this model gives a poor description of the experimental $T_c(p)$ values. We connected the extended charge transfer model to the pressure induced changes of the compressibilities. With this ansatz the calculated $T_c(p)$ values show an excellent agreement with the experimentally determined values.*

A distinctive feature of the HTSCs is the dependence of the superconducting properties on the charge carrier concentration $n_h$ in the $CuO_2$ planes. Beyond a particular $n_h$ value $T_c$ increases from zero to a maximum transition temperature $T_{c,max}$ at an optimum hole doping $n_{h,opt}$ and decreases again in the overdoped region. Presland et al.[1] showed that the parabolic shape of $T_c(n_h)$ is a universal relation if $T_c/T_{c,max}$ is plotted versus $n_h$ with $n_{h,opt}$=0.16 holes per $CuO_2$ plane:

$$\frac{T_c}{T_{c,max}} = 1 - \left(\frac{n_h - n_{h,opt}}{0.11}\right)^2 . \quad (1)$$

The hole concentration $n_h$ can be changed by chemical doping, i.e. for $YBa_2Cu_3O_x$ (Y123) by the variation of the oxygen content x or by substitution of $Ca^{2+}$ for $Y^{3+}$ or $La^{3+}$ for $Ba^{2+}$ [2,3,4]. The resulting $n_h$ values are difficult to determine from the nominal chemical composition due to the lack of knowledge of the true position of the doped atoms, mutual influence of the doping atoms (e.g. Ca influences the oxygen content [2]) and problems of estimating the portion of holes within the unit cell that really enters the $CuO_2$ planes and contributes to $n_h$ [5]. Another method to influence $n_h$ is the application of pressure which gives the advantage to preserve the chemical composition of the sample. Pressure redistributes the holes within the unit cell resulting in a positive $dn_h/dp$ [6]. The experiments led to an extension of this charge transfer model (CTM) by considering a pressure dependence of $T_{c,max}$ because at optimal doping there still exists a positive pressure effect $dT_{c,max}/dp$ [8]. This extended CTM is often used to describe the results of $T_c(p)$.

At higher pressure, $T_c(p)$ shows deviations from this model. For example $T_c(p)$ of Tl- and Hg-superconductors is too high for higher pressures, which is assigned to intrinsic properties of inequivalent $CuO_2$ planes [9,10].

At lower pressures it was shown that the extended CTM is valid for overdoped $Y_{1-y}Ca_yBa_2Cu_3O_x$ [11]. From these results we obtain $dT_{c,max}/dp = 0.8$ K/GPa and a charge transfer of $dn_h/dp = 3.7 \cdot 10^{-3}$ GPa$^{-1}$ independent of the Ca content. In the present experiment we have investigated $Y_{0.8}Ca_{0.2}Ba_2Cu_3O_{6.92}$ samples. The high oxygen content allows to neglect oxygen ordering effects within the CuO-chain subsystem [12,13] and the high Ca content ensures that the sample is overdoped. We determined $T_c(p)$ and lattice parameters to check the validity of the extended charge transfer model for this sample in the extended pressure range up to 10 GPa.

The $Y_{0.8}Ca_{0.2}Ba_2Cu_3O_{6.92}$ samples were grown in Ca-stabilized $ZrO_2$ crucibles. Details of the sample preparation may be found in Ref. 14. From EDX and neutron-scattering experiments the occupation of Ba sites by Ca is known to be less than 2%. The experiments were performed in a diamond-anvil cell using NaF as pressure transmitting medium and pressure gauge. A special coil system surrounding one of the diamonds allows susceptibility measurements, and with a x-ray system we could take spectra to determine lattice parameters. Fig. 1 shows the real part of the susceptibility of $Y_{0.8}Ca_{0.2}Ba_2Cu_3O_{6.92}$ as a function of temperature. $T_c$ has been determined from a 10% criterion as shown in Fig. 1. The transition at 65.3 K was obtained at ambient pressure. With increasing pressure the transition shifts to lower temperatures and is nearly conserving the shape as is to be expected for an overdoped sample under quasi hydrostatic pressure conditions. Fig. 2 shows the obtained $T_c(p)$ values for our $Y_{0.8}Ca_{0.2}Ba_2Cu_3O_{6.92}$ samples as a function of pressure.

To apply the CTM we have to calculate the $n_h$ value for our samples at ambient pressure. Due to the problems mentioned in the introduction this is not done by the nominal chemical composition, but we follow the method of

**Structure and $T_c$ of $Y_{1-y}Ca_yBa_2Cu_3O_x$ under High Pressure**

Obertelli et al.[15] and use Eq. (1) to calculate $n_h$ at ambient pressure from $T_c$ and $T_{c,max}$. To obtain the $T_{c,max}$ we measured $T_c(p=0)$ at numerous oxygen contents resulting in $T_{c,max}$ = 82.6 K.

With these values we obtain $n_h|_{p=0}$ = 0.21 for $Y_{0.8}Ca_{0.2}Ba_2Cu_3O_{6.92}$.

From earlier experiments under He pressures up to 0.6 GPa, we took $dT_{c,max}/dp$ = 0.8 K/GPa and a pressure induced charge transfer $dn_h/dp = 3.7 \cdot 10^{-3}$ GPa$^{-1}$ [11]. From this we obtained in the extended CTM $T_{c,max}(p) = T_{c,max}|_{p=0} + p \cdot 0.8$ K/GPa and $n_h(p) = n_h|_{p=0} + p \cdot 3.7 \cdot 10^{-3}$ GPa$^{-1}$. $T_c(p)$ is calculated from these values via Eq.(1) giving the dashed curve shown in Fig. 2. The dotted curve in Fig 2. represents $T_c(p)$ from the simple CTM with a constant $T_{c,max}(p)$ = 82.6 K for comparison. The bad description of the simple CTM is improved by the extended CTM due to the pressure dependent $T_{c,max}(p)$, but both models fail for $p > 4$ GPa.

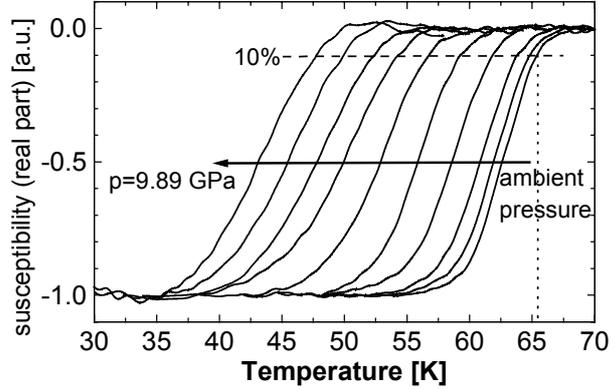

Fig.1. Real part of susceptibility of $Y_{0.8}Ca_{0.2}Ba_2Cu_3O_{6.92}$ as a function of temperature at several pressures.

In early investigations it has been suggested that $T_c(p)$ should be converted to a volume dependence [16]. However, when $V(p)$ has been used to follow this suggestion usually the bulk modulus $B_0$ was used. As changes in the pressure dependent bulk modulus $B(p) = B_0 + B' \cdot p$ are not included in $B_0$, with such a description the volume is proportional to pressure and the pressure dependence of $T_c$ is not changed.

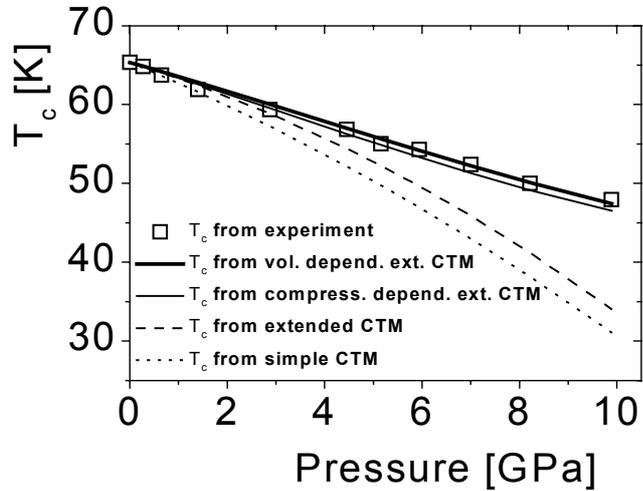

Fig. 2. $T_c$ of $Y_{0.8}Ca_{0.2}Ba_2Cu_3O_{6.92}$ as obtained from the experiment (□). The lines represent calculations from charge transfer models (CTM) as explained in the text.

On the other hand it is known that high-$T_c$ superconductors have an unusual large pressure dependence $B'$ which is neglected by using $B_0$ only [17]. This large $B'$ is caused by a hard matrix containing a soft element e.g. the Apex - CuO chain - Apex element between the adjacent $CuO_2$ planes in Y123 which causes the enhanced c-axis compressibility. A model compound to demonstrate such an enhanced $B'$ caused by soft structural elements is $C_{60}$ with soft van-der-Waals bindings between extreme hard spheres [18].

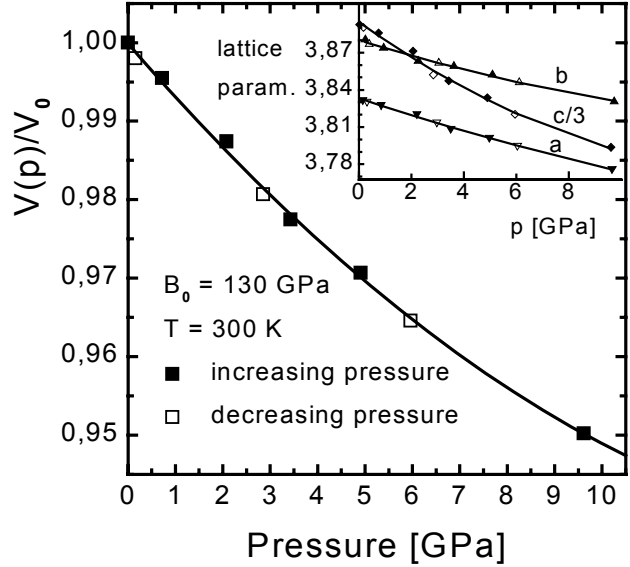

Fig. 3. Pressure dependence of the reduced volume of $Y_{0.9}Ca_{0.1}Ba_2Cu_3O_{6.93}$. The inset gives the *a*-, *b*- and *c*/3-lattice parameter as a function of pressure.

Therefore, we tried to consider the pressure dependence of $V/V_0$ for explaining our $T_c(p)$ data. As we obtained in earlier experiments almost the same $V(p)/V_0$ dependence for a Ca content of 0% or 10% we use for this discussion the values of Y123 with 10% Ca. The lattice parameters are shown in the inset of Fig. 3. The results confirm the expected high compressibility in c-axis direction and the large change of the linear c-axis compressibility with increasing pressure, which is expressed by the curvature of $c/3(p)$. The $V(p)/V_0$ dependence from the data is shown in Fig.3.

The simplest way to convert $T_c(p)$ to a $T_c(V(p))$ dependence is to describe $n_h(p)$ and $T_{c,max}(p)$ not as linear in *p*, but to be linear in the relative volume decrease which is $1-V(p)/V_0$. Therefore

$$n_h(p) = n_h|_{V_0} + \eta \cdot (1-V(p)/V_0) \quad (3)$$

$$T_{c,max}(p) = T_{c,max}|_{V_0} + \tau \cdot (1-V(p)/V_0) \quad (4)$$

At $V = V_0$ the new charge transfer coefficient $\eta$ is coupled to $dn_h/dp$ and the bulk modulus by

$$\eta = \frac{dn_h}{d(1-V/V_0)} = \frac{dn_h}{dp} \cdot \frac{dp}{d(1-V/V_0)} = \frac{dn_h}{dp} \cdot B_0 \quad (5)$$

# Structure and $T_c$ of $Y_{1-y}Ca_yBa_2Cu_3O_x$ under High Pressure

At $V = V_0$ $\tau$ is coupled in the same way to $dT_{c,max}/dp$ via $\tau = dT_{c,max}/dp \cdot B_0$. With the experimental values $dn_h/dp = 3.7 \cdot 10^{-3}$ GPa$^{-1}$, $dT_{c,max}/dp = 0.8$ K/GPa and $B_0 = 130$ GPa we obtain $\eta = 0.481$ and $\tau = 104$ K. Using Eq. (3) and (4), the $V(p)/V_0$ dependence shown in Fig.3 and the extended CTM, we obtain the $T_c(V(p))$ dependence plotted in Fig. 2 as a thick solid line. There is an excellent agreement of the resulting curve with the experimental values.

From uniaxial pressure data it is known that c-axis pressure is mainly responsible for $dn_h/dp$, and pressure in $a$-, $b$-axis direction influences $T_{c,max}$[4,19]. Therefore we couple $dn_h/dp$ and $dT_{c,max}/dp$ with the pressure dependent $c$-axis and $a$-,$b$-axis compressibilities, respectively, in the same way as explained above for $V(p)/V_0$. The resulting curve is plotted as a thin solid line in Fig. 2, showing an excellent agreement with the experimental points, too.

Our results on $Y_{0.8}Ca_{0.2}Ba_2Cu_3O_{6.92}$ support strongly the idea that the compressibility change under pressure has to be considered as well for the change in $T_{c,max}$ as for pressure induced charge transfer. This may partially explain the deviations between the experimentally observed $T_c(p)$ at higher pressures[9,10] and the extended CTM which uses constant $dT_{c,max}/dp$ and $dn_h/dp$. However, it has to be examined for Tl- and Hg-superconductors if the pressure dependence of the compressibility has similar influence on charge transfer and $dT_{c,max}/dp$.